\documentclass[review]{elsarticle}

\usepackage{tikz}
\usepackage{verbatim}
\usepackage{textcomp}
\usepackage{algpseudocode}
\usepackage{mdframed}
\usepackage[normalem]{ulem}
\usepackage{epstopdf}
\usepackage{subcaption}
\usepackage{graphicx}

\usetikzlibrary{backgrounds, calc,trees,positioning,arrows,chains,shapes.geometric, decorations.pathreplacing,decorations.pathmorphing,shapes,patterns, matrix,shapes.symbols,fit,scopes,external}

\usepackage{pgfplots,pgfplotstable}
\pgfplotsset{compat=newest}

\usepackage{booktabs} 
\usepackage{balance}
\usepackage{listings}
\usepackage{xcolor} 
\usepackage{soul}

\usepackage{amsmath,amssymb,amsfonts}
\usepackage{lipsum}
\usepackage{array,multirow}
\usepackage{siunitx}
\usepackage[acronyms,nonumberlist,nopostdot,nomain,nogroupskip]{glossaries}
\usepackage{lineno,hyperref}

\modulolinenumbers[5]

\journal{Simulation Modelling Practice and Theory}

\definecolor{seablue}{RGB}{05,102,141}
\definecolor{alabamacrimson}{RGB}{172,00,54}
\definecolor{darkgreen}{RGB}{03,49,46}
\definecolor{metallicsunburst}{RGB}{168,118,62}
\definecolor{silver}{RGB}{193,193,193}

\long\def\/*#1*/{}

\newacronym{quic}{QUIC}{Quick UDP Internet Connections}
\newacronym{3gpp}{3GPP}{3rd Generation Partnership Project}
\newacronym{adc}{ADC}{Analog to Digital Converter}
\newacronym{5g}{5G}{5th generation}
\newacronym{aimd}{AIMD}{Additive Increase Multiplicative Decrease}
\newacronym{am}{AM}{Acknowledged Mode}
\newacronym{amc}{AMC}{Adaptive Modulation and Coding}
\newacronym{aqm}{AQM}{Active Queue Management}
\newacronym{awgn}{AGWN}{Additive White Gaussian Noise}
\newacronym{balia}{BALIA}{Balanced Link Adaptation}
\newacronym{bdp}{BDP}{Bandwidth-Delay Product}
\newacronym{bf}{BF}{Beamforming}
\newacronym{cc}{CC}{Congestion Control}
\newacronym{cdf}{CDF}{Cumulative Distribution Function}
\newacronym{cn}{CN}{Core Network}
\newacronym{cqi}{CQI}{Channel Quality Information}
\newacronym{cp}{CP}{Control Plane}
\newacronym{csirs}{CSI-RS}{Channel State Information - Reference Signal}
\newacronym{dc}{DC}{Dual Connectivity}
\newacronym{dce}{DCE}{Direct Code Execution}
\newacronym{dci}{DCI}{Downlink Control Information}
\newacronym{dl}{DL}{Downlink}
\newacronym{dmr}{DMR}{Deadline Miss Ratio}
\newacronym{dmrs}{DMRS}{DeModulation Reference Signal}
\newacronym{e2e}{E2E}{End-to-End}
\newacronym{ecn}{ECN}{Explicit Congestion Notification}
\newacronym{edf}{EDF}{Earliest Deadline First}
\newacronym{enb}{eNB}{evolved Node Base}
\newacronym{epc}{EPC}{Evolved Packet Core}
\newacronym{es}{ES}{Edge Server}
\newacronym{fdma}{FDMA}{Frequency Division Multiple Access}
\newacronym{fdd}{FDD}{Frequency Division Duplexing}
\newacronym[firstplural=Radio Access Technologies (RATs)]{rat}{RAT}{Radio Access Technology}
\newacronym{fs}{FS}{Fast Switching}
\newacronym{ftp}{FTP}{File Transfer Protocol}
\newacronym{gnb}{gNB}{Next Generation Node Base}
\newacronym{harq}{HARQ}{Hybrid Automatic Repeat reQuest}
\newacronym{hetnet}{HetNet}{Heterogeneous Network}
\newacronym{hh}{HH}{Hard Handover}
\newacronym{hol}{HOL}{Head-of-Line}
\newacronym{ia}{IA}{Initial Access}
\newacronym{imt}{IMT}{International Mobile Telecommunication}
\newacronym{iot}{IoT}{Internet of Things}
\newacronym{ipsd}{IPSD}{Interference Power Spectral Density}
\newacronym{kpi}{KPI}{Key Performance Indicator}
\newacronym{kpis}{KPIs}{Key Performance Indicators}
\newacronym{los}{LOS}{Line of Sight}
\newacronym{lte}{LTE}{Long Term Evolution}
\newacronym{m2m}{M2M}{Machine to Machine}
\newacronym{mac}{MAC}{Medium Access Control}
\newacronym{mc}{MC}{Multi-Connectivity}
\newacronym{mcs}{MCS}{Modulation and Coding Scheme}
\newacronym{mec}{MEC}{Mobile Edge Cloud}
\newacronym{mi}{MI}{Mutual Information}
\newacronym{mimo}{MIMO}{Multiple-Input Multiple-Output}
\newacronym{mmwave}{mmWave}{millimeter wave}
\newacronym{mr}{MR}{Maximum Rate}
\newacronym{mss}{MSS}{Maximum Segment Size}
\newacronym{mtd}{MTD}{Machine-Type Device}
\newacronym{mtu}{MTU}{Maximum Transmission Unit}
\newacronym{nfv}{NFV}{Network Function Virtualization}
\newacronym{nlos}{NLOS}{Non Line of Sight}
\newacronym{nr}{NR}{New Radio}
\newacronym{ofdm}{OFDM}{Orthogonal Frequency Division Multiplexing}
\newacronym{o2i}{O2I}{Outdoor-To-Indoor}
\newacronym{pdcch}{PDCCH}{Physical Downlonk Control Channel}
\newacronym{pdcp}{PDCP}{Packet Data Convergence Protocol}
\newacronym{pdsch}{PDSCH}{Physical Downlink Shared Channel}
\newacronym{pdu}{PDU}{Packet Data Unit}
\newacronym{pf}{PF}{Proportional Fair}
\newacronym{pgw}{PGW}{Packet Gateway}
\newacronym{phy}{PHY}{Physical}
\newacronym{pbch}{PBCH}{Physical Broadcast Channel}
\newacronym[plural=\gls{mme}s,firstplural=Mobility Management Entities (MMEs)]{mme}{MME}{Mobility Management Entity}
\newacronym{prb}{PRB}{Physical Resource Block}
\newacronym{pss}{PSS}{Primary Synchronization Signal}
\newacronym{pucch}{PUCCH}{Physical Uplink Control Channel}
\newacronym{pusch}{PUSCH}{Physical Uplink Shared Channel}
\newacronym{rach}{RACH}{Random Access Channel}
\newacronym{ran}{RAN}{Radio Access Network}
\newacronym{red}{RED}{Random Early Detection}
\newacronym{rem}{REM}{Radio Environment Map}
\newacronym{rems}{REMs}{Radio Environment Maps}
\newacronym{rf}{RF}{Radio Frequency}
\newacronym{rlc}{RLC}{Radio Link Control}
\newacronym{rlf}{RLF}{Radio Link Failure}
\newacronym{rrc}{RRC}{Radio Resource Control}
\newacronym{rrm}{RRM}{Radio Resource Management}
\newacronym{rr}{RR}{Round Robin}
\newacronym{rs}{RS}{Remote Server}
\newacronym{rsrp}{RSRP}{Reference Signal Received Power}
\newacronym{rss}{RSS}{Received Signal Strength}
\newacronym{rtt}{RTT}{Round Trip Time}
\newacronym{rv}{RV}{Reduncy Version}
\newacronym{rw}{RW}{Receive Window}
\newacronym{rx}{RX}{Receiver}
\newacronym{sa}{SA}{standalone}
\newacronym{sack}{SACK}{Selective Acknowledgment}
\newacronym{sap}{SAP}{Service Access Point}
\newacronym{sch}{SCH}{Secondary Cell Handover}
\newacronym{scm}{SCM}{Spatial Channel Model}
\newacronym{scoot}{SCOOT}{Split Cycle Offset Optimization Technique}
\newacronym{sdma}{SDMA}{Spatial Division Multiple Access}
\newacronym{sinr}{SINR}{Signal to Interference plus Noise Ratio}
\newacronym{sm}{SM}{Saturation Mode}
\newacronym{snr}{SNR}{Signal to Noise Ratio}
\newacronym{son}{SON}{Self-Organizing Network}
\newacronym{ss}{SS}{Synchronization Signal}
\newacronym{srs}{SRS}{Sounding Reference Signal}
\newacronym{sss}{SSS}{Secondary Synchronization Signal}
\newacronym{tb}{TB}{Transport Block}
\newacronym{tcp}{TCP}{Transmission Control Protocol}
\newacronym{tdd}{TDD}{Time Division Duplexing}
\newacronym{tdma}{TDMA}{Time Division Multiple Access}
\newacronym{tfl}{TfL}{Transport for London}
\newacronym{tm}{TM}{Transparent Mode}
\newacronym{trp}{TRP}{Transmitter Receiver Pair}
\newacronym{tti}{TTI}{Transmission Time Interval}
\newacronym{ttt}{TTT}{Time-to-Trigger}
\newacronym{tx}{TX}{Transmitter}
\newacronym{ue}{UE}{User Equipment}
\newacronym{ul}{UL}{Uplink}
\newacronym{uml}{UML}{Unified Modeling Language}
\newacronym{um}{UM}{Unacknowledged Mode}
\newacronym{utc}{UTC}{Urban Traffic Control}
\newacronym{vm}{VM}{Virtual Machine}
\newacronym{rsrq}{RSRQ}{Reference Signal Received Quality}
\newacronym{rssi}{RSSI}{Received Signal Strength Indicator}
\newacronym{crs}{CRS}{Cell Reference Signal}
\newacronym{comp}{CoMP}{Coordinated Multi-Point}
\newacronym{cran}{C-RAN}{Cloud \acrlong{ran}}
\newacronym{ca}{CA}{Carrier Aggregation}
\newacronym{cco}{CC}{Carrier Component}
\newacronym{nsa}{NSA}{Non Stand Alone}
\newacronym{embb}{eMBB}{Enhanced Mobility Broadband}
\newacronym{bsr}{BSR}{Buffer Status Report}
\newacronym{srb}{SRB}{Service Radio Bearer}
\newacronym{sctp}{SCTP}{Stream Control Transmission Protocol}
\newacronym{mptcp}{MPTCP}{Multi-path TCP}
\newacronym{ietf}{IETF}{Internet Engineering Task Force}
\newacronym{os}{OS}{Operating System}
\newacronym{tls}{TLS}{Transport Layer Security}
\newacronym{rfc}{RFC}{Request for Comments}
\newacronym{http}{HTTP}{HyperText Transfer Protocol}
\newacronym{nat}{NAT}{Network Address Translation}
\newacronym{api}{API}{Application Programming Interface}
\newacronym{rto}{RTO}{Retransmission Timeout}
\newacronym{psc}{PSC}{Public Safety Communication}
\newacronym{rpgm}{RPGM}{Reference Point Group Mobility}
\newacronym{ic}{IC}{Incident Command}
\newacronym{rsu}{RSU}{Road Side Unit}
\newacronym{uav}{UAV}{Unmanned Aerial Vehicle}
\newacronym{iab}{IAB}{Integrated Access and Backhaul}
\newacronym{psd}{PSD}{Power Spectral Density}
\newacronym{mpc}{MPC}{Multi Path Component}
\newacronym{rt}{RT}{Ray Tracer}
\newacronym{aoa}{AoA}{Angle of Arrival}
\newacronym{aod}{AoD}{Angle of Departure}
\newacronym{inr}{INR}{Interference to Noise Ratio}
\newacronym{qd}{QD}{Quasi Deterministic}
\newacronym{wlan}{WLAN}{Wireless Local Area Network}
\newacronym{cad}{CAD}{Computer-aided Design}
\newacronym{ap}{AP}{Access Point}
\newacronym{sta}{STA}{Station}
\newacronym{nrmse}{NRMSE}{Normalized Root Mean Square Error}
\newacronym{ut}{UT}{User Terminal}
\newacronym{bs}{BS}{Base Station}

\begin{document}
\begin{frontmatter}

\title{Calibration of the 5G-LENA System Level Simulator in 3GPP reference scenarios}

\author{Katerina Koutlia}
\ead{katerina.koutlia@cttc.es}

\author{Biljana Bojovic}
\ead{biljana.bojovic@cttc.es}

\author{Zoraze Ali}
\ead{zoraze.ali@cttc.es}

\author{Sandra Lag\'en}
\ead{sandra.lagen@cttc.es}

\address{Centre Tecnol\`ogic de Telecomunicacions de Catalunya (CTTC/CERCA) \\
Avinguda Carl Friedrich Gauss, 7 \\
08860 Castelldefels, Barcelona, Spain \\
\{kkoutlia, bbojovic, zali, slagen\}@cttc.es}

\begin{abstract}
Due to the rapid technology evolution and standardization activity in the mobile communication networks, there is the need for the research community to be able to develop, test and evaluate new and/or already existing solutions before industrial or real-network implementation. As such, it is essential to have an open-source tool that provides an alternative solution to that of industrial proprietary simulators that are not available for public usage. ns-3 5G-LENA simulator is an end-to-end open-source NR system-level simulator that allows extensive research to be performed. However, it is of great importance to guarantee that the results obtained using the simulator can be comparable to that of industrial simulators and real networks. For this reason, calibrating the simulator based on 3GPP defined specifications is crucial. Based on the above, in this paper we calibrate the ns-3 5G-LENA simulator according to the 3GPP reference results for NR-based outdoor deployments. Moreover, we explore the REM feature provided by the simulator, to ease the calibration process and understand better the radio environment. Results show the resemblance of the simulator performance to that of simulators used as references by 3GPP.
\end{abstract}


\end{frontmatter}

\section{Introduction}

Fifth Generation New Radio (5G NR) has been introduced by 3GPP with Release 15~\cite{38300}, promising to offer low-latency, high-reliable, and high-speed communications, as well as extended coverage. Among other features, NR supports operation in both sub-6 GHz (FR1) and millimeter-wave (FR2) bands~\cite{parkvall:17}, with the latter enabling the support of high data rates with the usage of advanced beamforming techniques and wide-bandwidth operations~\cite{pi:11}.
5G networks are designed in such a way so that they would employ novel and evolved architectures and technologies~\cite{Peng16}. As such, rapid evaluation, implementation, and standardization of these architectures and technologies can play a key role in order to fulfil the challenges 5G networks face and meet the demands imposed by the users in terms of quality of service and performance.

In order to evaluate and improve already existing features or deploy new ones, it is of great importance for the research community to have the means to perform testing and validations. Simulation environments are the enablers for such steps to be carried out, with final target the actual deployment. In contrast to the industrial proprietary simulators (even those used by companies in 3GPP) that are not available for public usage, and whose results are thus not reproducible, open-source 5G simulators are available to the public. The open access to the source code allows the development of new features, the modification of already available ones, and the testing and evaluation of them. Examples of open-source 5G simulators are given in Table~\ref{table:5GSimulators}~\cite{electronics9030468}.

\begin{table}[!t]
\small
\centering
\caption{5G open-source simulators~\cite{electronics9030468}}
\begin{tabular}{|p{4cm}|p{7cm}|}
 \hline
 \textbf{Simulator} & \textbf{Main Characteristics} \\
 \hline
 NYUSIM~\cite{NYUSIM} & Link-level simulator for statistical channel modeling, simulation code with an easy-to-use interface and carrier frequencies from 2 to 73 GHz \\
 \hline
 Vienna 5G Simulators~\cite{vienna5g} & Link-level and system-level simulators, for large scale simulations (including hundreds of nodes) and supporting parallelization \\
 \hline
 WISE~\cite{wiseSim} & System-level simulator for multi-tier orientations \\
 \hline
 GTEC 5G Simulator~\cite{GTECsim} & Link-level simulator \\
 \hline
 5G Toolbox by Matlab~\cite{matlab5g} & Link-level simulator, focused on channel modeling and signal generation \\
 \hline
 Simu5G~\cite{simu5G} & OMNeT++-based System-level, end-to-end simulator \\
 \hline
 5G-air-simulator~\cite{5GairSim} & System-level, end-to-end simulator for modeling the 5G air interface \\
 \hline
 ns–3 mmWave~\cite{nyummwave} & System-level, full protocol stack, end-to-end simulator of 3GPP NR at millimeter-wave bands \\
 \hline
 5G-LENA~\cite{PATRICIELLO2019101933} & System-level, full protocol stack, end-to-end  simulator of 3GPP NR, supporting 0.5-100 GHz frequency bands \\
 \hline
\end{tabular}
\label{table:5GSimulators}
\end{table}

Moreover, there is the necessity for the research community to perform evaluations and obtain results that are analogous to that of the 3GPP standardization and the industrial simulators. 
Notably, simulators used by companies in 3GPP are required to pass through a calibration procedure, so that, even if they are not public, at least they are aligned with similar deployments, propagation conditions, and performances.
On that respect, ITU-R has defined in the IMT-2020 report~\cite{IMT-2020} certain specifications and technical performance requirements to be fulfilled by the candidate radio interface technologies. All in all, it is essential that any simulation tool, before exploiting it for evaluation and validation, is calibrated to the reference results in IMT-2020 and 3GPP simple reference scenarios, guaranteeing the validity of the results with respect to the 3GPP specifications. 

For the calibration process, it can also result helpful to exploit \glspl{rem}, if available, in order to verify the deployment and propagation conditions and eventually find the proper configuration. The \gls{rem} tool can provide information related to the network interference and the signal quality, providing better knowledge of the radio environment~\cite{rem3, rem2}. This information can be very useful for the study and analysis of the radio propagation environment, and eventually can be used for the calibration process.

In this work, we focus on the 5G-LENA module of the ns-3~\cite{ns3} discrete-event network simulator. 5G-LENA~\cite{PATRICIELLO2019101933} is an NR system-level simulator openly available to the research community~\cite{5gLena}, that derives from the ns-3 LENA~\cite{BaldoLena} and ns-3 mmWave~\cite{nyummwave} modules. 5G-LENA implements a high-fidelity full protocol stack and allows cross-layer evaluations and the study of end-to-end (E2E) performances.

The ns-3 LENA simulator of Fourth Generation Long Term Evolution (4G LTE) technology was calibrated in~\cite{marinescu} to a set of 3GPP reference scenarios defined in 3GPP TR 36.814, including Urban Macro (UMa) and Rural Macro (RMa) setups, in terms of downlink per user average Signal-to-Interference-plus-Noise ratio (SINR) and normalized user throughput. In~\cite{PATRICIELLO2019101933}, we calibrated the ns-3 5G-LENA simulator in Indoor Hotspot scenario setups, according to 3GPP TR 38.802, in terms of SINR and Signal-to-Noise ratio (SNR). However, as of today, the calibration of ns-3 5G-LENA simulator in outdoor (and mixed outdoor/indoor) deployments is missing.

In this respect, this paper is focused on the calibration of the 5G-LENA simulator based on the 3GPP reference results for outdoor deployments. In addition, since the 5G-LENA accounts with the REM feature \cite{rem3}, the calibration process followed here has been carried out by exploiting these maps, proving their importance in the network design and analysis. Based on the above, the main contribution of this paper is to give to the community an open source tool that allows the testing, evaluation, validation, and experimentation of existing and/or new features, guaranteeing the resemblance of the results to that of an industrial private product or of a real network.

The rest of the paper is organized as follows. Section~\ref{sec:TestEnvironments} describes the different test environments defined for calibration by ITU, while the detailed simulation settings and the parameters for the selected subset of scenarios under which we have calibrated the 5G-LENA simulator, are presented in Section~\ref{sec:Scenarios}. In Section~\ref{sec:5gLena} we give a brief description of the required changes and extensions performed in the 5G-LENA module in order to achieve the calibration of it according to the IMT-2020 Report~\cite{IMT-2020}. The performance evaluation against the 3GPP considered industrial simulators, as well as the REM evaluation are given in Section~\ref{sec:Results}. Finally, the most important conclusions are drawn in Section~\ref{sec:Conclusions}.

\section{Calibration Test Environments for 5G NR}
\label{sec:TestEnvironments}
ITU has defined in Report ITU-R IMT-2020 \cite{IMT-2020} a number of test environments and evaluation configurations for the calibration of the candidate system-level simulators. Also, it provides the reference results to calibrate with. In particular, it defines three different usage scenarios (enhanced mobile broadband (eMBB), massive machine type communications (mMTC) and ultra-reliable and low latency communications (URLLC)) and combines each of them with one or several geographic environment(s), leading to five different test environments \cite{IMT-2020}, namely:
\begin{itemize}
    \item Indoor Hotspot-eMBB
    \item Dense Urban-eMBB
    \item Rural-eMBB
    \item Urban Macro–mMTC
    \item Urban Macro–URLLC
\end{itemize}
Furthermore, for each test environment, different evaluation configurations are defined \cite{IMT-2020}. These different evaluation configurations basically vary some parameters (like the carrier frequency, the total transmit power, the simulation bandwidth, the number of antenna elements per gNB/UE, and the gNB/UE noise figure) for a given test environment. Indoor Hotspot-eMBB and Rural-eMBB include three configurations (Configuration A, Configuration B, and configuration C), while Dense Urban-eMBB, Urban Macro–mMTC and Urban Macro–mMTC consider just two different configurations (Configuration A and Configuration B). Further details with respect to the technical performance requirements, the service and spectrum aspect requirements, and the evaluation guidelines can be found in~\cite{M.2410, M.2411, M.2412}, respectively. The results obtained by companies in 3GPP for the scenarios defined in~\cite{M.2412}, have been summarized and compiled in~\cite{RP180524}.

\section{Considered Reference Scenarios}
\label{sec:Scenarios}
Indoor scenarios have already been used for the system-level calibration of the 5G-LENA simulator in previous work~\cite{PATRICIELLO2019101933}. However, outdoor scenarios have not been considered up to this point. As such, in this work we focus on the outdoor scenarios with eMBB, i.e., Rural-eMBB and Dense Urban-eMBB test environment scenarios, defined in~\cite{IMT-2020}.
For the Rural-eMBB, we have focused on Configuration A and Configuration B, while for the Dense Urban-eMBB we have focused on Configuration A. Other non-eMBB scenarios and configurations are left for future work, because calibration is a time consuming process and requires big efforts for each configuration.

Depending on the scenario under evaluation, UEs might be indoor/outdoor, while the antenna configuration and height, among other parameters, of both gNBs and UEs, are also varied.
Moreover, let us notice that all scenarios have been evaluated under full buffer traffic, as indicated by 3GPP reference results. 
The specific details related to the calibration settings of these scenarios are given in Sections~\ref{sec:Rural} and~\ref{sec:Um}, respectively.

\subsection{Rural-eMBB} 
\label{sec:Rural}

Rural Configuration A and Rural Configuration B are defined with carrier frequencies of 700 MHz and 4 GHz, respectively, focusing on FR1 band. The test environment follows a tri-sectorial hexagonal topology as described in Section~\ref{sec:NetTop}. 10 UEs per sector are randomly deployed following a uniform distribution, where 50\% are considered as indoor and 50\% as outdoor. UE applications are configured to start at the same time in order to achieve full buffer traffic since the beginning of the applications start time. However, note that the moment at which a UE enters the network (UE connection) is random because of the random access procedure and RRC, but applications are configured to start after the RRC connection is established.
The detailed calibration parameters for the Rural configurations A and B are given in~\cite{RP180524}. We present a summary of them, according to which we have also configured 5G-LENA simulator for calibration, in Table~\ref{table:Ruraltable}.

\begin{table}[!t]
\small
\centering
\caption{Rural-eMBB Configuration A and B Parameters}
\begin{tabular}{|p{4.0cm}|p{2.7cm}|p{2.7cm}|}
 \hline
 \textbf{Parameter} & \textbf{Configuration A} & \textbf{Configuration B} \\
 \hline
 Carrier frequency & 700 MHz & 4 GHz \\
 \hline
 & \multicolumn{2}{|c|}{\textbf{Common Parameters}} \\
 \hline
 Bandwidth & \multicolumn{2}{|c|}{10MHz} \\
 \hline
 Inter-site distance & \multicolumn{2}{|c|}{1732 m} \\
 \hline
 Sectors & \multicolumn{2}{|c|}{30/150/270 degrees} \\
 \hline
 BS antenna height & \multicolumn{2}{|c|}{35 m} \\
 \hline
 UE antenna height & \multicolumn{2}{|c|}{1.5 m} \\
 \hline
 BS transmit power & \multicolumn{2}{|c|}{46 dBm/sector} \\
  \hline
 UE transmit power & \multicolumn{2}{|c|}{23 dBm} \\
 \hline
 UEs deployment & \multicolumn{2}{|c|}{10 UEs per sector, randomly dropped} \\
 \hline
 Propagation model & \multicolumn{2}{|c|}{3GPP RMa TR 38.901} \\
 \hline
 BS antenna array & \multicolumn{2}{|c|}{1 TXRU: 8x1 (3GPP elements)} \\
 \hline
 BS horizontal/vertical antenna separation & \multicolumn{2}{|c|}{0.5/0.8} \\
 \hline
 BS antenna element gain & \multicolumn{2}{|c|}{8 dBi} \\
 \hline
 UE antenna array & \multicolumn{2}{|c|}{1 TXRU: 1x1 (isotropic element)} \\
 \hline
 UE antenna element gain & \multicolumn{2}{|c|}{0 dBi} \\
 \hline
 BS noise figure & \multicolumn{2}{|c|}{5 dB} \\
 \hline
 UE noise figure & \multicolumn{2}{|c|}{7 dB} \\
 \hline
 Noise power spectral density & \multicolumn{2}{|c|}{-174 dBm/Hz} \\
 \hline
 Percentage of high/low loss building type & \multicolumn{2}{|c|}{100\% low loss (channel model B)} \\
 \hline
 Device deployment & \multicolumn{2}{|c|}{50\% indoor, 50\% outdoor } \\
 \hline
 UE Mobility Model & \multicolumn{2}{|c|}{Fixed speed 3 km/h for all UEs} \\
 \hline
 Minimum BS-UE distance & \multicolumn{2}{|c|}{10 m} \\
 \hline
 Inter-Site Interference Modelling & \multicolumn{2}{|c|}{Explicitly modelled} \\
 \hline
 Downtilt angle & \multicolumn{2}{|c|}{90\si{\degree}} \\
 \hline
 Traffic Model & \multicolumn{2}{|c|}{Full Buffer} \\
 \hline
\end{tabular}
\label{table:Ruraltable}
\end{table}

For the calculation of Pathloss, the TR 38.901 model \cite{TR38901} is used as given in the following equation:
\begin{equation}
    \text{PL} = \text{PL}_{b} + \text{PL}_{tw} + \text{PL}_{in} + N(0, \sigma^2_\text{P}), \label{Pathloss}
\end{equation}
where $\text{PL}_{b}$ is the basic outdoor Pathloss given in Table 7.4.1-1 of TR 38.901 \cite{TR38901} according to the LOS and NLOS conditions, $\text{PL}_{tw}$ is the \gls{o2i} building penetration loss, $\text{PL}_{in}$ is the inside loss dependent on the depth into the building, and $\sigma_\text{P}$ is the standard deviation for the penetration loss. $\text{PL}_{tw}$, $\text{PL}_{in}$ and $\sigma_{P}$ are given in Table 7.4.3-2 of TR 38.901~\cite{TR38901}. As indicated also in Table~\ref{table:Ruraltable}, for both Configuration scenarios of the Rural-eMBB, the Outdoor-To-Indoor penetration losses are defined as 100\% low loss, meaning that all the indoor UEs will experience low penetration losses.

For the \gls{scm} and antenna model focusing on wireless
channels between 0.5 and 100 GHz, we follow the TR 38.901~\cite{TR38901} specifications, as described in~\cite{tommaso:20}, where the implemented channel model for the ns-3 simulator is presented in detail. However, we needed to extend the 3GPP \gls{scm} implemented in~\cite{tommaso:20} to account for the \gls{o2i} penetration losses and dual-polarized antennas (more details are given in Section~\ref{sec:5gLena}). 

\subsection{Dense Urban-eMBB}
\label{sec:Um}
The Dense Urban-eMBB scenario consists of a macro and a micro layer. However, in 3GPP for the calibration purpose, only the macro layer is considered. As such, the considered topology here is focused on the macro layer. The network deployment follows the same tri-sectorial hexagonal topology as described in Section~\ref{sec:NetTop}, where 10 UEs per sector are randomly and uniformly deployed. The Pathloss model, the SCM and the antenna model used for the calibration process are as described in Section~\ref{sec:Rural}. Notice that in this scenario, the deployed UEs are 20\% outdoor and 80\% indoor, while the percentage of high loss and low loss building type is set to 20\% high loss and 80\% low loss. This means that the experienced penetration losses for an indoor UE in this scenario can be 20\% high and 80\% low. Moreover, let us point out that in the Dense Urban-eMBB, the antenna height of the UEs follows a 3D deployment option: i.e., it is varied depending on the type; outdoor UEs have $h_\text{UT}$ = 1.5 m, while $h_\text{UT}$ (in m) of indoor UEs is assigned randomly based on the following formula:
\begin{equation}
    h_\text{UT} = 3(\text{nfl} - 1) + 1.5,
    \label{randomHUt}
\end{equation}
where nfl $\sim$ uniform(1, Nfl) and Nfl $\sim$ uniform(4, 8).

The rest of the calibration parameters, as also used to calibrate 5G-LENA simulator, are given in Table~\ref{table:DenseTable}.

\begin{table}[!t]
\small
\centering
\caption{Dense Urban-eMBB Configuration A Parameters}
\begin{tabular}{|p{4cm}|p{6cm}|}
 \hline
 \textbf{Parameter} & \textbf{Configuration A} \\
 \hline
 Carrier frequency & 4 GHz \\
 \hline
 Bandwidth & 10MHz \\
 \hline
 Inter-site distance & 200 m \\
 \hline
 Sectors & 30/150/270 degrees \\
 \hline
 BS antenna height & 25 m \\
 \hline
 UE antenna height & outdoor UEs: 1.5 m, indoor UEs: random based on equation~\eqref{randomHUt} \\
 \hline
 BS transmit power & 41 dBm/sector \\
  \hline
 UE transmit power & 23 dBm \\
 \hline
 UEs deployment & 10 UEs per sector, randomly dropped \\
 \hline
 Propagation model & 3GPP RMa TR 38.901 \\
 \hline
 BS antenna array & 1 TXRU: 4x8 (3GPP elements) \\
 \hline
 BS horizontal/vertical antenna separation & 0.5/0.8 \\
 \hline
 BS antenna element gain & 8 dBi \\
 \hline
 UE antenna array & 1 TXRU: 1x1 (isotropic element) \\
 \hline
 UE antenna element gain & 0 dBi \\
 \hline
 BS noise figure & 5 dB \\
 \hline
 UE noise figure & 7 dB \\
 \hline
 Noise power spectral density & -174 dBm/Hz \\
 \hline
 Percentage of high/low loss building type & 80\% low loss - 20\% high loss \\
 \hline
 Device deployment & 80\% indoor, 20\% outdoor \\
 \hline
 UE Mobility Model & Fixed speed 3 km/h for all UEs \\
 \hline
 Minimum BS-UE distance & 10 m \\
 \hline
 Inter-Site Interference Modelling & Explicitly modelled \\
 \hline
 Downtilt angle & 90\si{\degree} \\
 \hline
 BeamForming Method & BeamSearch with specific azimuth/zenith direction angles for analog beam steering\\
 \hline
 Traffic Model & Full Buffer \\
 \hline
\end{tabular}
\label{table:DenseTable}
\end{table}

\subsection{Network Topology}
\label{sec:NetTop}

\begin{figure*}[t!]
\includegraphics[width=1.0 \columnwidth]{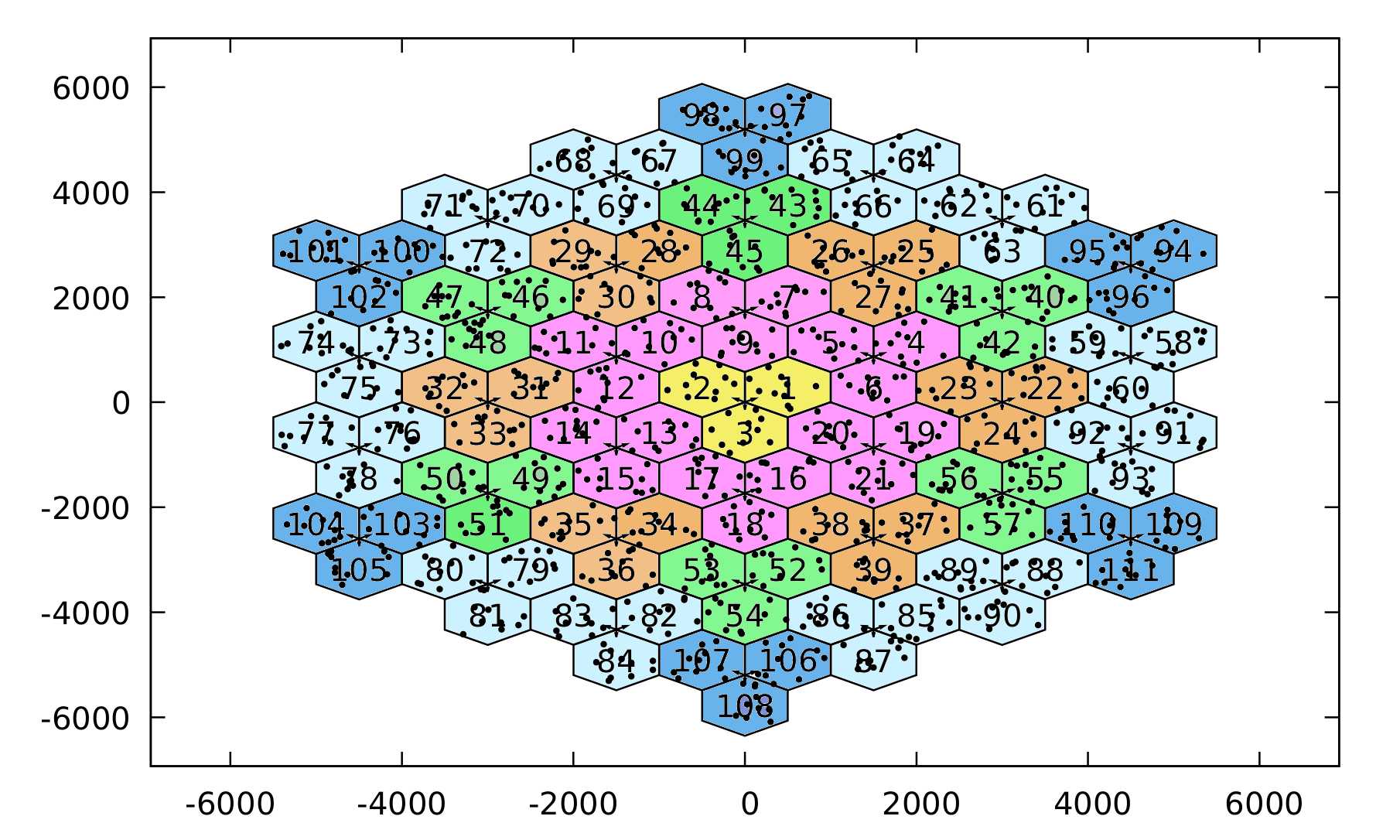}
\centering
\caption{Wrap-around hexagonal deployment. The rings are colored in the following way: 0th in yellow, 1st in light purple, 2nd in orange, 3rd in green, 4th in light blue, and 5th in dark blue.}
\label{fig:wrap-around}
\end{figure*}

The network layout consists in a hexagonal topology with 37 sites of 3 sectors each, thus leading to 111 Base Stations (BS), as shown in Figure \ref{fig:wrap-around}. However, in the measurements we consider only the 21 inner BSs, while the 111 BSs are simulated to account for the wrap-around effect~\cite{M.2412}, as described in detail in Section~\ref{sec:5gLena}. Moreover, let us point out that in order to speed up the simulations, we ignore channel realizations and interferences between nodes located at a distance larger than 2 times the inter-site distance.

\subsection{Evaluation - KPIs}

The \glspl{kpi} used for the evaluation of the candidate radio interface technologies are the Coupling Gain and the Downlink Geometry. According to \cite{Nomor}, the Coupling Gain includes the pathloss, the antenna gains and the average fast fading, while it excludes processing gains such as the ones obtained by beamforming techniques, except for analog beamforming gains of the transmission/reception units (TXRUs) where applicable. In particular, the Coupling Gain can be computed as follows:
\begin{equation}
    \text{Coupling Gain [dB]}= P^T - P^R
\end{equation}
where $P^T$ is the transmitted power and $P^R$ is the received power.

The Downlink Geometry on the other hand, is described in \cite{Nomor} as the ratio of received signal power to the sum of interference and noise power where all signals are averaged individually over the used bandwidth. Similarly, to the Coupling Gain, the Downlink Geometry does not include any processing gain at transmitter or receiver except with analog beamforming where applicable. As such, the Downlink Geometry can be considered as a wideband SINR, and it is measured per TXRU using the following equation:
\begin{equation}
    \text{SINR}= \frac{P^R}{I+N} 
    \label{sinr} 
\end{equation}
where $P^R$ is the received power, $I$ is the sum of interference powers and $N$ is the noise power.

\section{5G-LENA Implementation}
\label{sec:5gLena}
The main key features of the 5G-LENA simulator are: 1)	Flexible to work from 400 MHz to 100 GHz, using the 3GPP spatial channel model TR 38.901, 2) Non-standalone architecture: 5G RAN and 4G EPC, 3) Flexible and automatic configuration of the NR frame structure through multiple numerologies, 4)	Orthogonal Frequency-Division Multiple Access (OFDMA)-based access with variable TTIs and Time-Division Multiple Access (TDMA)-based access, 5)	Restructuring and redesign of the MAC layer, including new flexible MACs schedulers that simultaneously consider time-and frequency-domain resources (resource blocks and symbols) both for TDMA and OFDMA-based access schemes with variable TTI, 6)	Uplink grant-based access scheme with scheduling request and 3GPP-compliant buffer status reporting, 7)	\gls{fdd} and \gls{tdd} supported, with flexible and configurable TDD patterns, 8)	NR-compliant processing timings (K1, K2), 9)	New Bandwidth Part (BWP) managers and the architecture to support operation through multiple BWPs and multiple Component Carriers (CCs), 10)	New NR-compliant physical layer abstraction, considering LDPC codes for data channels, and MCS tables 1 and 2 (including up to 256-QAM), 11)	Compatibility with 3GPP uniform planar arrays, with directional and isotropic radiation, 12)	Beamforming support, with various analog beamforming methods, ideal and realistic beamforming with SRS-based channel estimates, 13)	Dual-polarized MIMO support  with rank adaptation algorithms.

For the calibration process, we have further extended the 5G-LENA simulator with some additional specific features that were not supported previously and are needed for calibration. These features are described briefly in the following:
\begin{itemize}
    \item \textbf{3GPP SCM O2I Losses.} We have extended the SCM to include the possibility to add low and high penetration losses for a O2I channel condition, based on the TR 38.901 model \cite{TR38901} and as described in Section~\ref{sec:Rural}. For this, the \texttt{ChannelConditionModel} has been extended to calculate the O2I/O2O channel condition based on either the UE antenna height, or a random uniform variable set through the tested example. For the former case, the O2I condition is considered only for UEs with random antenna heights, while for the latter, O2I condition is assigned randomly based on the result of the random uniform variable. Moreover, in case the channel condition results to be O2I, low or high penetration losses are applied (as defined in Table 7.4.3-2 of TR 38.901~\cite{TR38901}) based on a threshold set in the user example. For this, the \texttt{ThreeGppPropagationLossModel} has also been extended to calculate and account for the O2I low/high penetration losses when calculating the final Pathloss. The O2I channel condition and low/high condition period updates are set equal to the LOS/NLOS channel condition update period.
    
    \item \textbf{Wrap-around}. To reproduce correctly the behavior of the network without introducing some network edge effects, it is necessary to simulate 
    an ``infinite`` cellular network. A famous technique to achieve that, is known as wrap-around. To support the wrap-around technique in this calibration study, we have extended the \texttt{HexagonalGridScenarioHelper} to allow the creation of the 4th and the 5th ring (previously, the maximum number of rings was 3). In this way, it is possible to simulate 5 rings.
    However, in the calibration results we consider only the measurements of the inner rings, i.e. those who are fully interfered, such as rings 0 and 1. In Figure~\ref{fig:wrap-around}, we illustrate the full simulation deployment. Each hexagon represents a single cell, and 3 hexagons with collocated transmitters form 1 site. Various definitions of what is a ring in a hexagonal deployment exist in the literature. In 5G-LENA, we adopted the definition according to which each ring is composed of the sites that are on the same distance from the center site (or 0th ring, shown in yellow in Figure~\ref{fig:wrap-around}). The different rings are denoted with different colors in Figure~\ref{fig:wrap-around}.

    \item \textbf{Support of 3D user deployment}. The \texttt{HexagonalGridScenarioHelper} has been further extended to create a deployment with UEs with mobility and UEs with 3D antenna height, where the speed and the percentage of UEs with random antenna height are specified through the tested example script. 
    
    \item \textbf{Dual-polarized antennas and beam sets}. We have extended the SCM to allow the simulation of dual-polarized antennas, at gNBs and UEs, based on the polarized antenna Model-2 in TR 38.901~\cite{TR38901}. Also, for the Dense Urban-eMBB scenario, we have implemented the beam sets at gNB/UEs according to~\cite{RP180524}, which specifies the directions for analog beam steering including specific azimuth and zenith angles.
 
    \item \textbf{Filter CQI=0.} Since 5G-LENA currently does not support Handover, and therefore due to the user mobility, there is the chance that users can be located in out-of-coverage areas, we omit from the SINR results calculations associated to a CQI (Channel Quality Indicator)=0. 
\end{itemize}

Notice that all these changes have been included in the 5G-LENA (nr module available at~\cite{ns3devCttc}), except the extension of the SCM to include O2I losses and the support of dual-polarized antennas that we implemented in the antenna and propagation modules of the ns-3-dev simulator (available at~\cite{ns3Calibration}), used by 5G-LENA.

Moreover, for the calibration purpose we have created a new example where we have defined the test environment scenarios. For each scenario, we have added a set of pre-defined parameters, meaning parameters that will not be changed during the simulations, such as the frequency or the bandwidth, while we have included some other parameters to be defined through the command line, in case we want to study variations in the KPIs based on different configurations. Examples include the gNB and UE polarization and slant angles, the beamforming method used, and the activation/deactivation of fading and shadowing.

Finally, for the Coupling Gain and SINR statistics, we have used the 5G-LENA traces already included in the module. We measure the SINR from the downlink data, assuming wideband-transmissions, no HARQ retransmissions, and single TXRU, as per its definition.

\section{Calibration Results}
\label{sec:Results}
In this section we present the simulation results, as obtained from the REM helper and from end-to-end (E2E) simulations. 

\begin{figure}[!t]
\includegraphics[width=1\columnwidth]{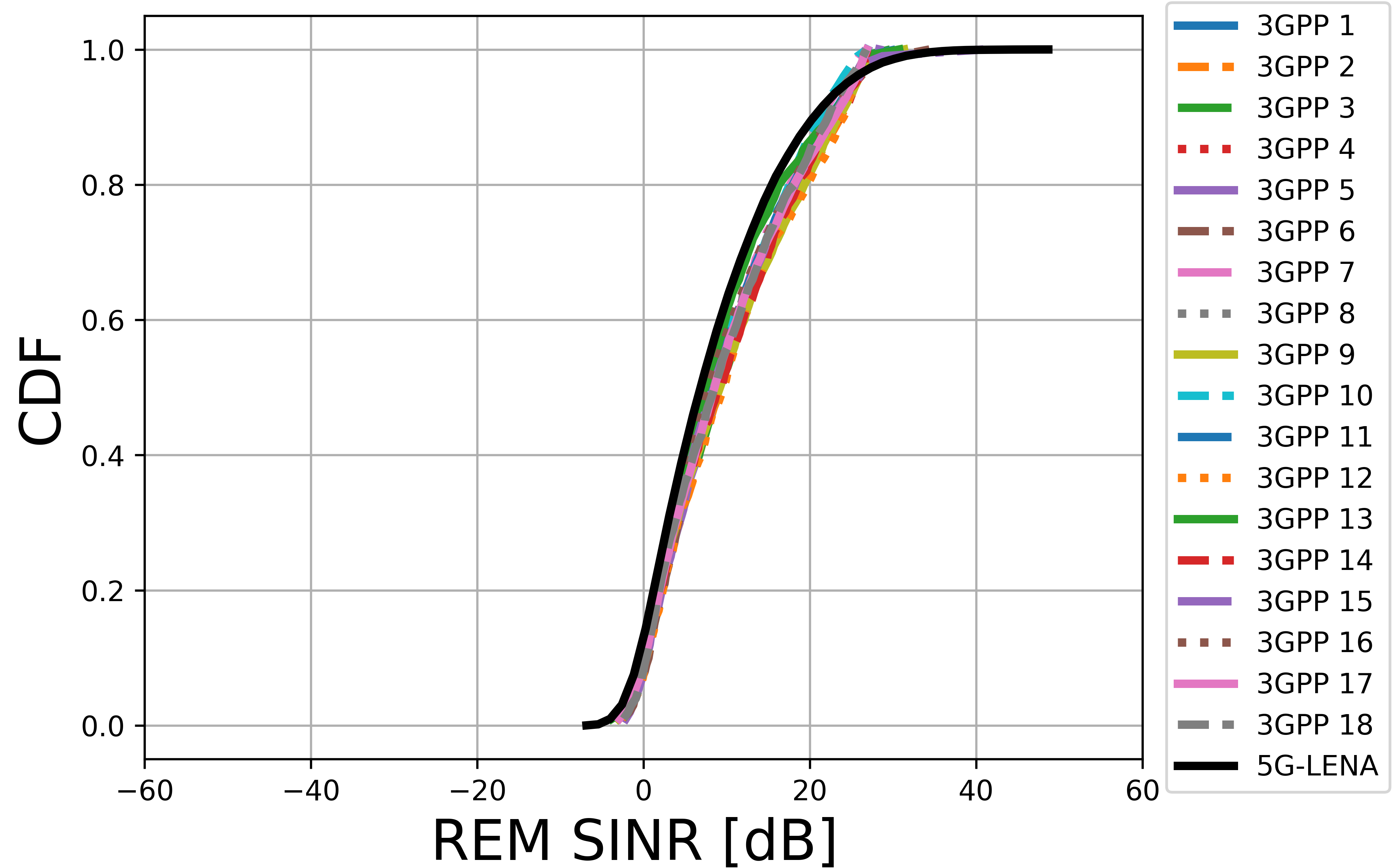}
\centering
\caption{Rural Configuration A - REM SINR CDF}
\label{fig:RuralARemSinr}
\end{figure}

\begin{figure}[!t]
\includegraphics[width=1\columnwidth]{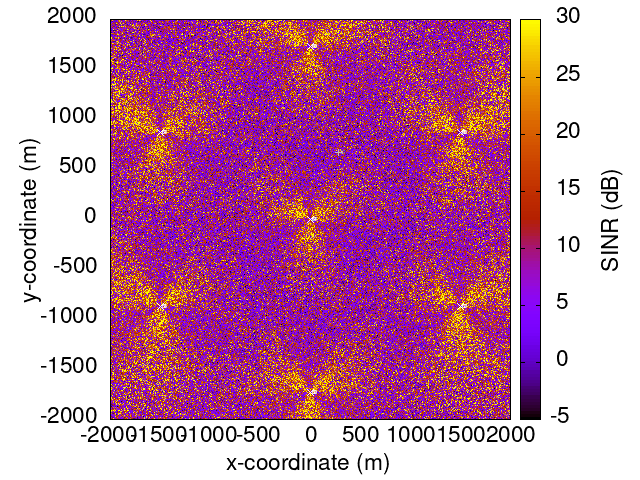}
\centering
\caption{Rural Configuration A - REM SINR map}
\label{fig:RuralARemMap}
\end{figure}

\subsection{Usage of REM for Validation}
5G-LENA's REM feature has been presented in~\cite{rem3}, implemented in the \texttt{NrRadioEnvironmentMapHelper} class. It allows the network status visualization, so that the phases of network design and analysis can be facilitated and accelerated. We have used the 5G-LENA REM feature as a means to understand the network conditions and find the proper configuration for the first studied case, the Rural-eMBB Configuration A and B test scenarios. Notice that REM implementation in 5G-LENA considers the best serving cell, but with worst-case interference conditions. The fact of considering the best serving cell is perfectly suited with 3GPP reference simulators and reference results, which assume an ideal handover with 0~dB handover margin (i.e., the strongest cell is selected) and UE attachment based on the best received power.

\begin{figure}[!t]
\includegraphics[width=1\columnwidth]{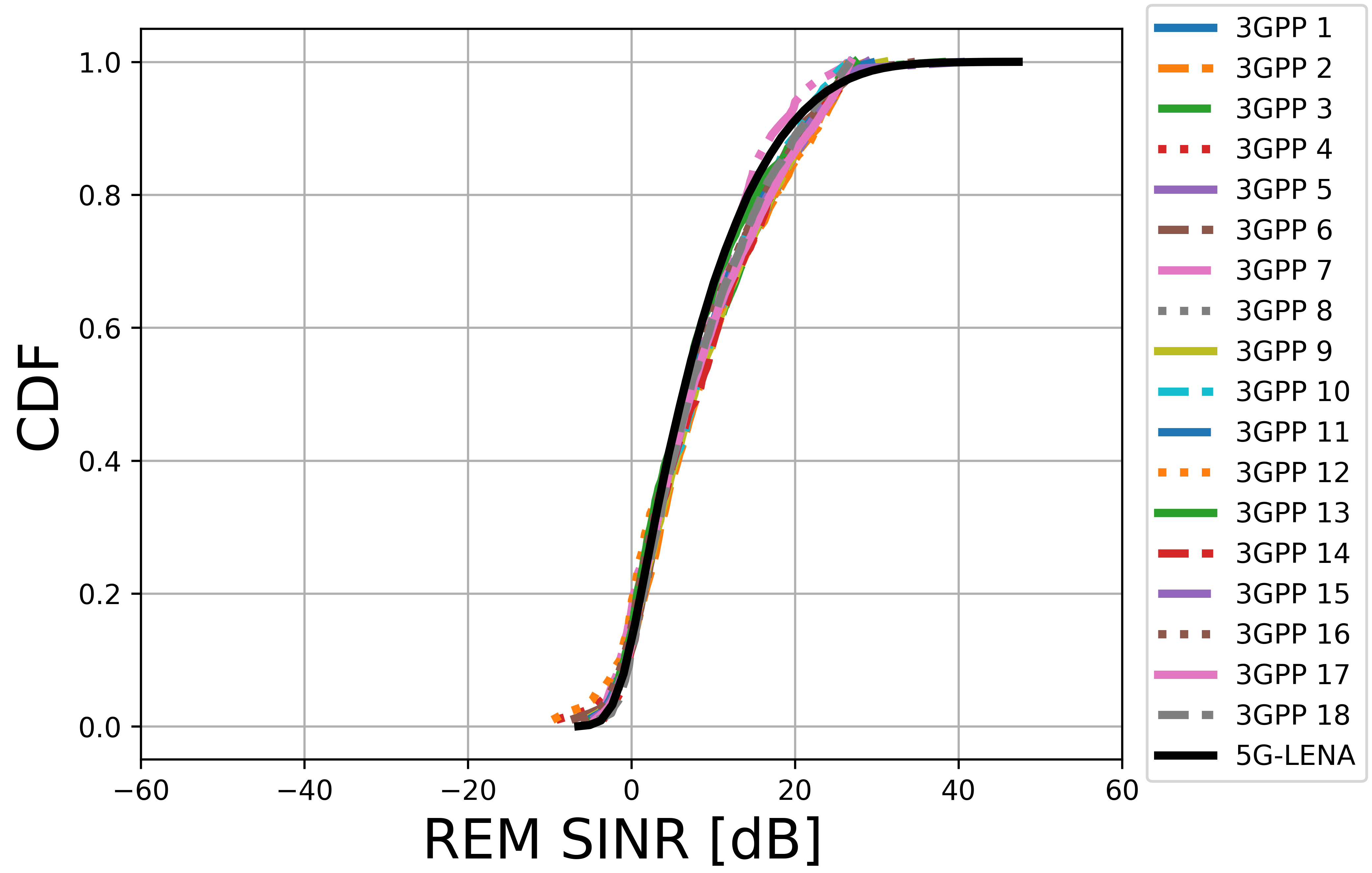}
\centering
\caption{Rural Configuration B - REM SINR CDF}
\label{fig:RuralBRemSinr}
\end{figure}

\begin{figure}[!t]
\includegraphics[width=1\columnwidth]{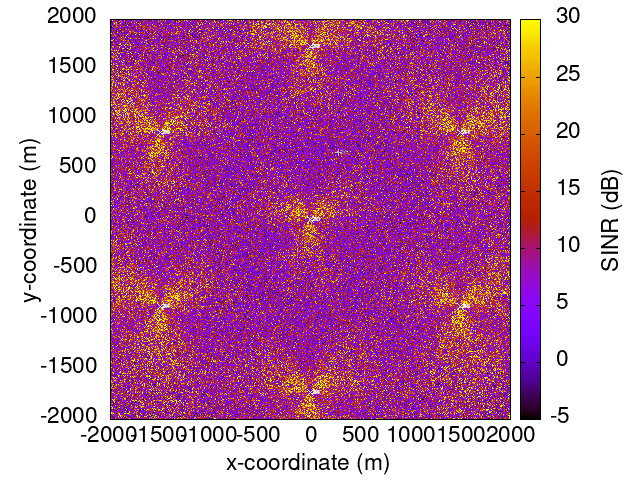}
\centering
\caption{Rural Configuration B - REM SINR map}
\label{fig:RuralBRemMap}
\end{figure}

The REM feature has been used in conjunction with the above mentioned implemented example. Results presented here are extracted from the SINR as measured at each REM point (each point of the map) and compared against the 3GPP reference simulators. Therefore, calibration results, focusing here on the SINR extracted using the 5G-LENA REM, are shown in Figures~\ref{fig:RuralARemSinr} and ~\ref{fig:RuralARemMap} for Configuration A and Figures~\ref{fig:RuralBRemSinr} and ~\ref{fig:RuralBRemMap} for Configuration B, respectively. Notice that Figures~\ref{fig:RuralARemSinr} and Figures~\ref{fig:RuralBRemSinr} depict the CDF for all the REM points, while Figures~\ref{fig:RuralARemMap} and ~\ref{fig:RuralBRemMap} depict the Beam Shape SINR REM map (Beam-Shape map is used to visualize the configuration of beamforming vectors, i.e., it shows the actual/effective coverage map in the tested scenario~\cite{rem3}).

Finally, let us point out that for the case of Dense Urban-eMBB scenario, we could not exploit the REM feature due to the fact that the random antenna UE height and consequently the penetration losses cannot be modeled since the \texttt{NrRadioEnvironmentMapHelper} is a tool that illustrates the network deployment in a 2D plane, and so cannot include the 3D UE deployment and its random antenna heights, rather a single REM point at a specific height is considered to show a 2D map. For this reason, we do not evaluate the Dense Urban-eMBB scenario with REM.

Let us also note that we have also extracted simulation results with 5 rings, under the Rural-eMBB scenarios, with both configurations (A and B). The results are not included here for the sake of brevity, but we obtained similar SINR CDFs with 5 simulated rings and filtering of the inner 0-1 rings, as the results that we show in this paper for 1 simulated ring. 

Based on the above, we can conclude that in the Rural scenarios, the SINR is well calibrated to 3GPP reference curves, which validates the configured deployment and propagation environment in the 5G-LENA example, as well as the implemented models in the 5G-LENA module.

\subsection{E2E Simulations}
Although REM SINR results are really useful for the calibration process, E2E simulations are necessary to confirm and validate the performance of the 5G-LENA simulator for both KPIs defined by 3GPP, i.e. Downlink Geometry and Coupling Gain. For this reason, we have performed system-level E2E simulations for both Rural-eMBB and Dense Urban-eMBB scenarios using the example described in Section~\ref{sec:5gLena}. 

\begin{figure}[!t]
\includegraphics[width=1\columnwidth]{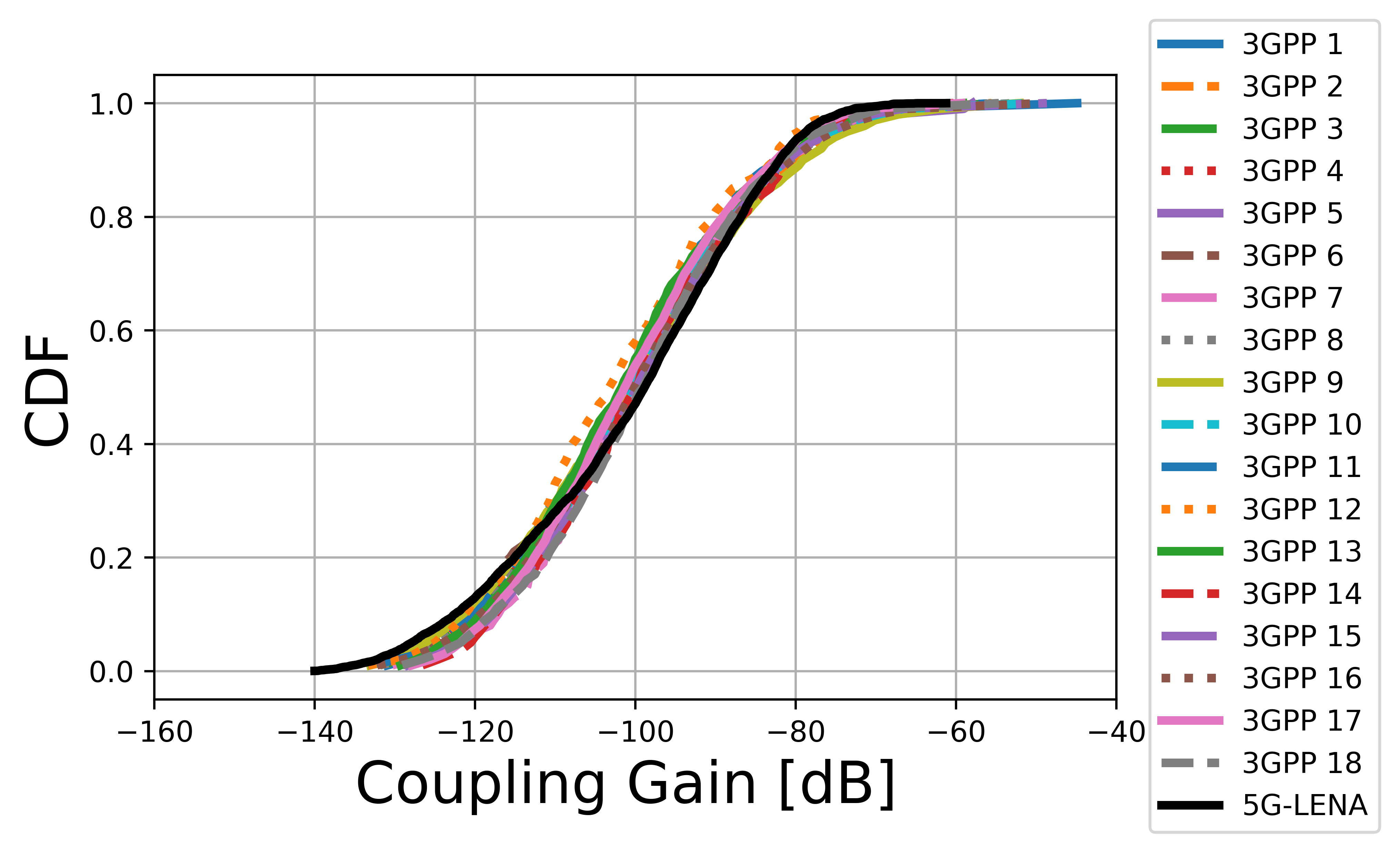}
\centering
\caption{Rural Configuration A - Coupling Gain}
\label{fig:CgRuralA}
\end{figure}

\begin{figure}[!t]
\includegraphics[width=1\columnwidth]{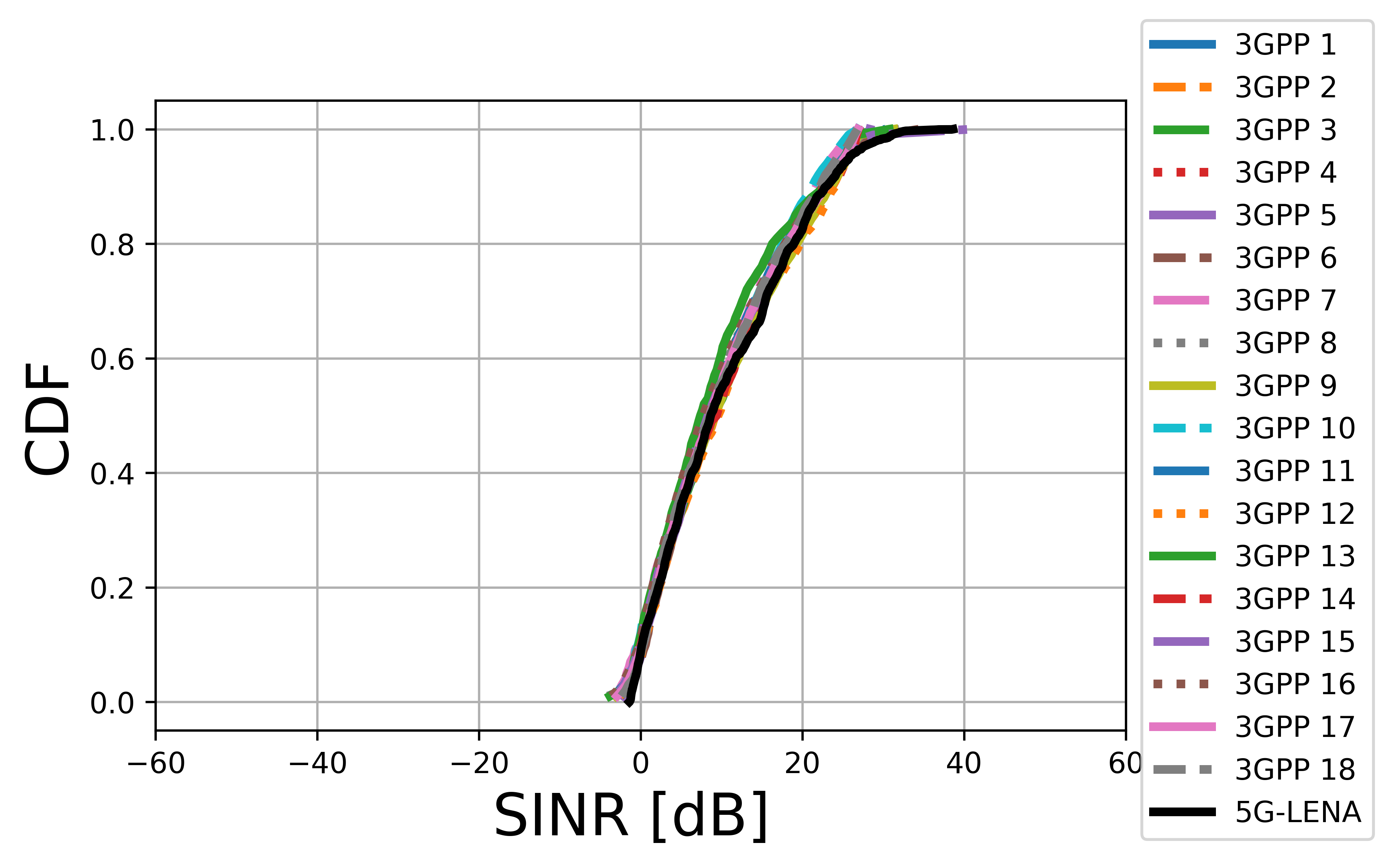}
\centering
\caption{Rural Configuration A - Downlink Geometry}
\label{fig:SinrRuralA}
\end{figure}

\begin{figure}[!t]
\includegraphics[width=1\columnwidth]{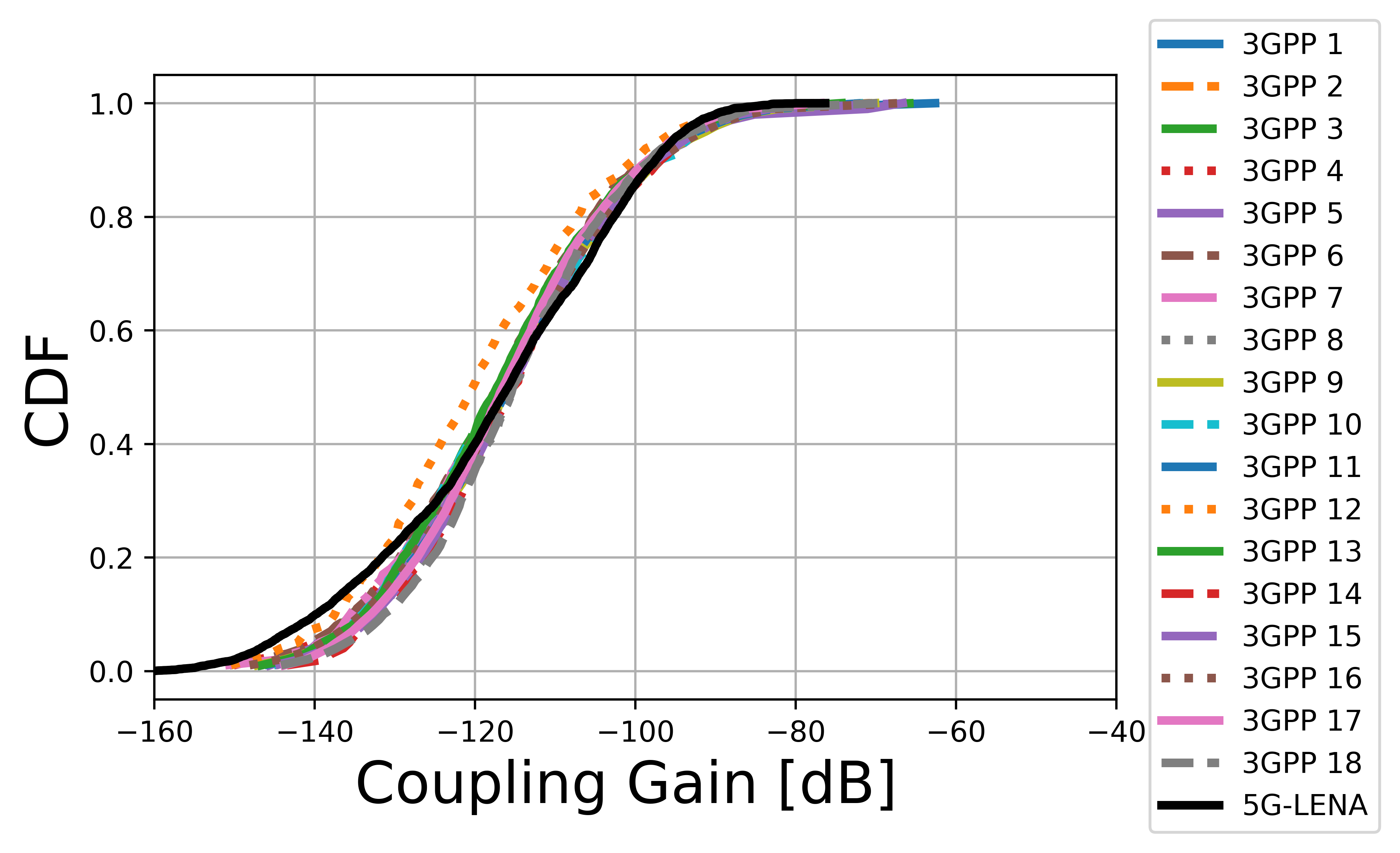}
\centering
\caption{Rural Configuration B - Coupling Gain}
\label{fig:CgRuralB}
\end{figure}

\begin{figure}[!t]
\includegraphics[width=1\columnwidth]{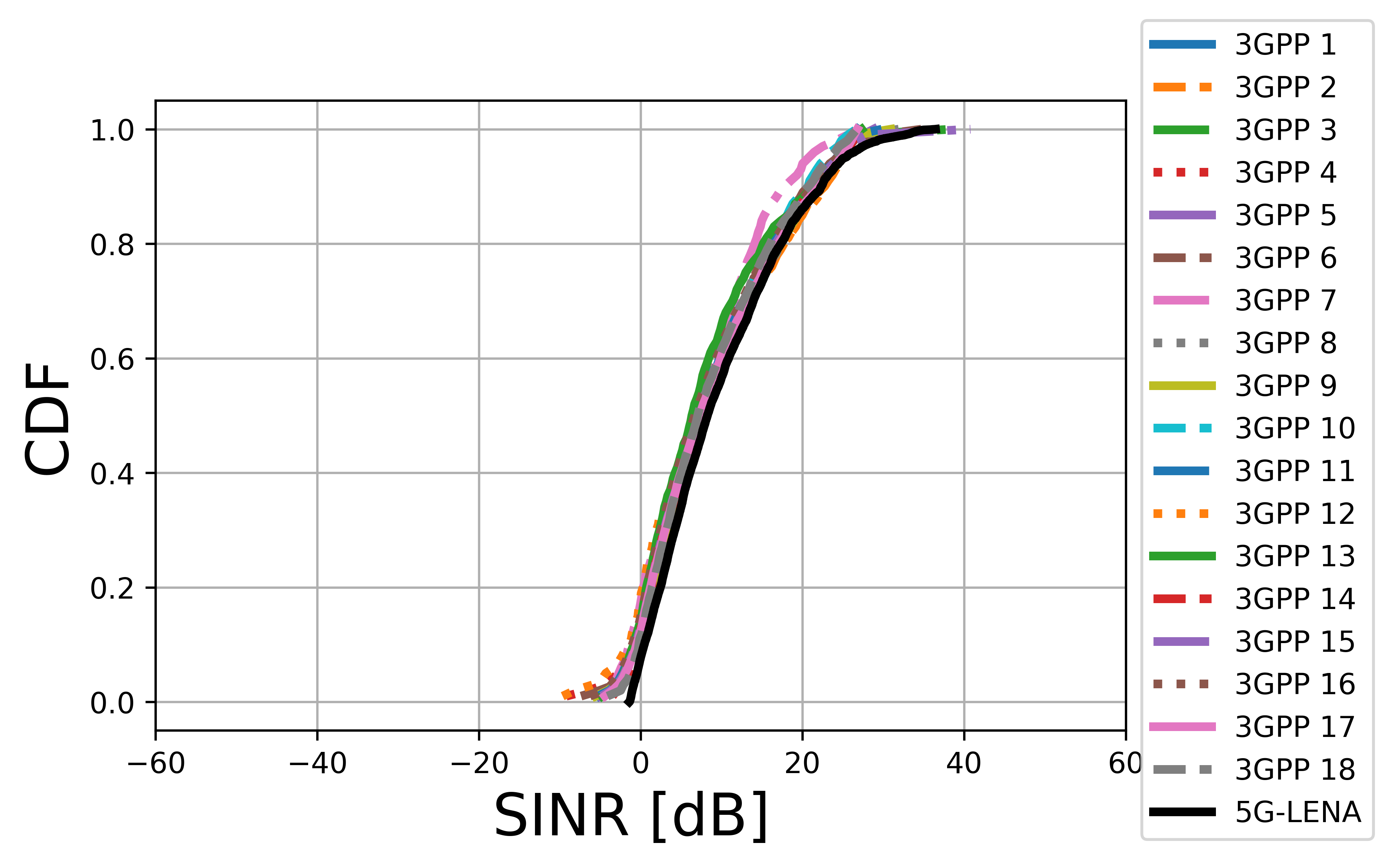}
\centering
\caption{Rural Configuration B - Downlink Geometry}
\label{fig:SinrRuralB}
\end{figure}

The first evaluated scenario is the Rural-eMBB under the two possible configurations (A and B), as described in Section~\ref{sec:Rural}. Results for the Coupling Gain are presented in Figure~\ref{fig:CgRuralA} for Configuration A, while for Configuration B are given in Figure~\ref{fig:CgRuralB}. Moreover, Figures~\ref{fig:SinrRuralA} and ~\ref{fig:SinrRuralB} depict the Downlink Geometry (Wide-band SINR) for Configurations A and B, respectively. Notice that the obtained results are compared against of that of various 3GPP industrial simulators provided in~\cite{RP180524}. As it can be observed from the figures, for both Rural configurations, the 5G-LENA simulator achieves very similar performance as the benchmark simulators with respect to both KPIs, with a slight difference only in the lower part of the Coupling Gain CDF and the SINR lower tail for Configuration B. Overall, it can be seen clearly that it meets the key technical performance requirements as defined in~\cite{M.2410}. Moreover, let us notice that even though 5G-LENA does not support (currently) handover and RSRP (Reference Signal Received Power) based attachment, these two features are modelled very well thanks to the assumptions we have taken under consideration, i.e., CQI0 filtering as described in Section~\ref{sec:5gLena} and attachment based on the closest cell. However, it would be of great interest to perform again the calibration of 5G-LENA once these two features are supported and compare the results against the 3GPP references and also against the current ones.

\begin{figure}[!t]
\includegraphics[width=1\columnwidth]{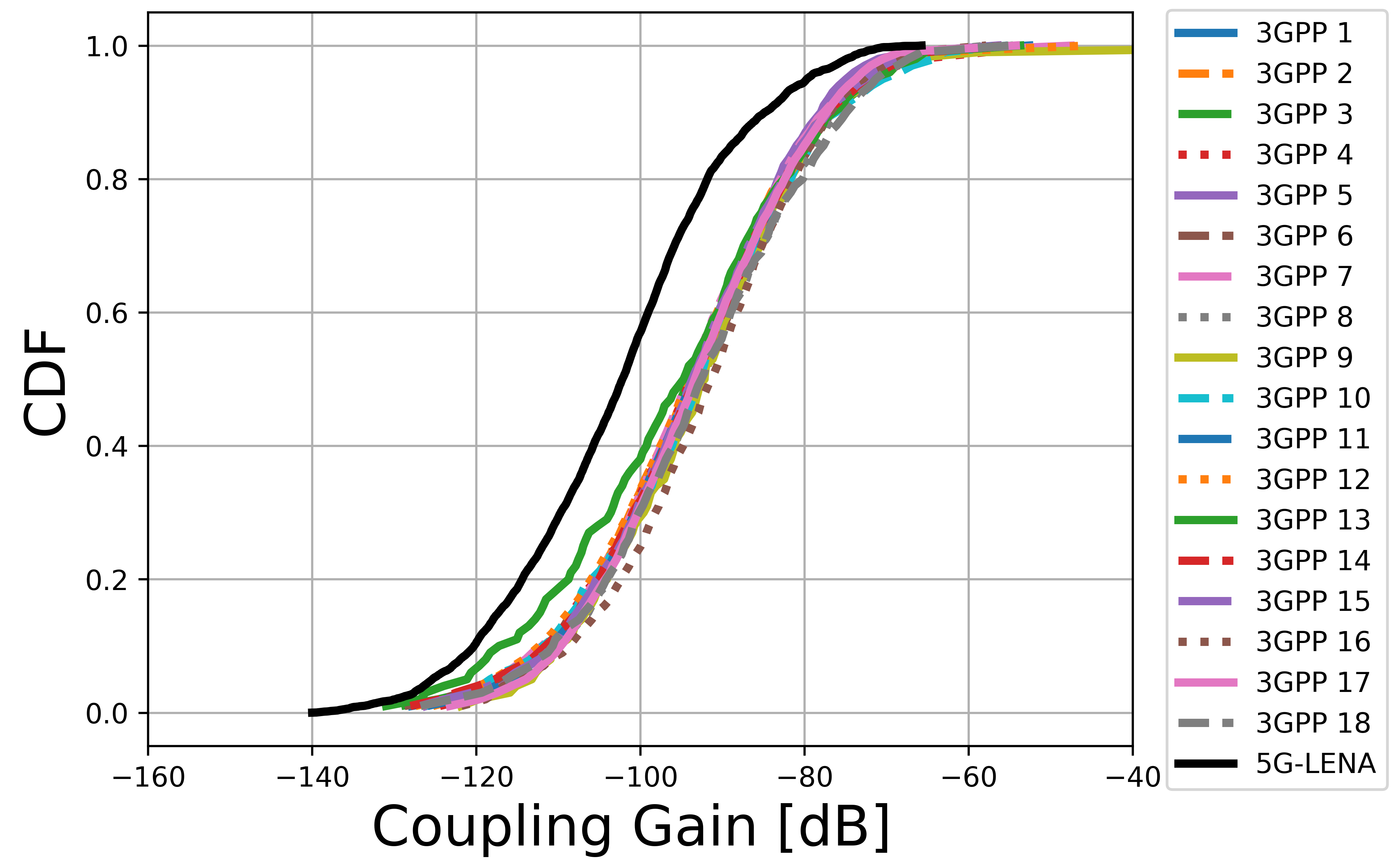}
\centering
\caption{Dense Urban Configuration A - Coupling Gain}
\label{fig:CgDenseA}
\end{figure}

\begin{figure}[!t]
\includegraphics[width=1\columnwidth]{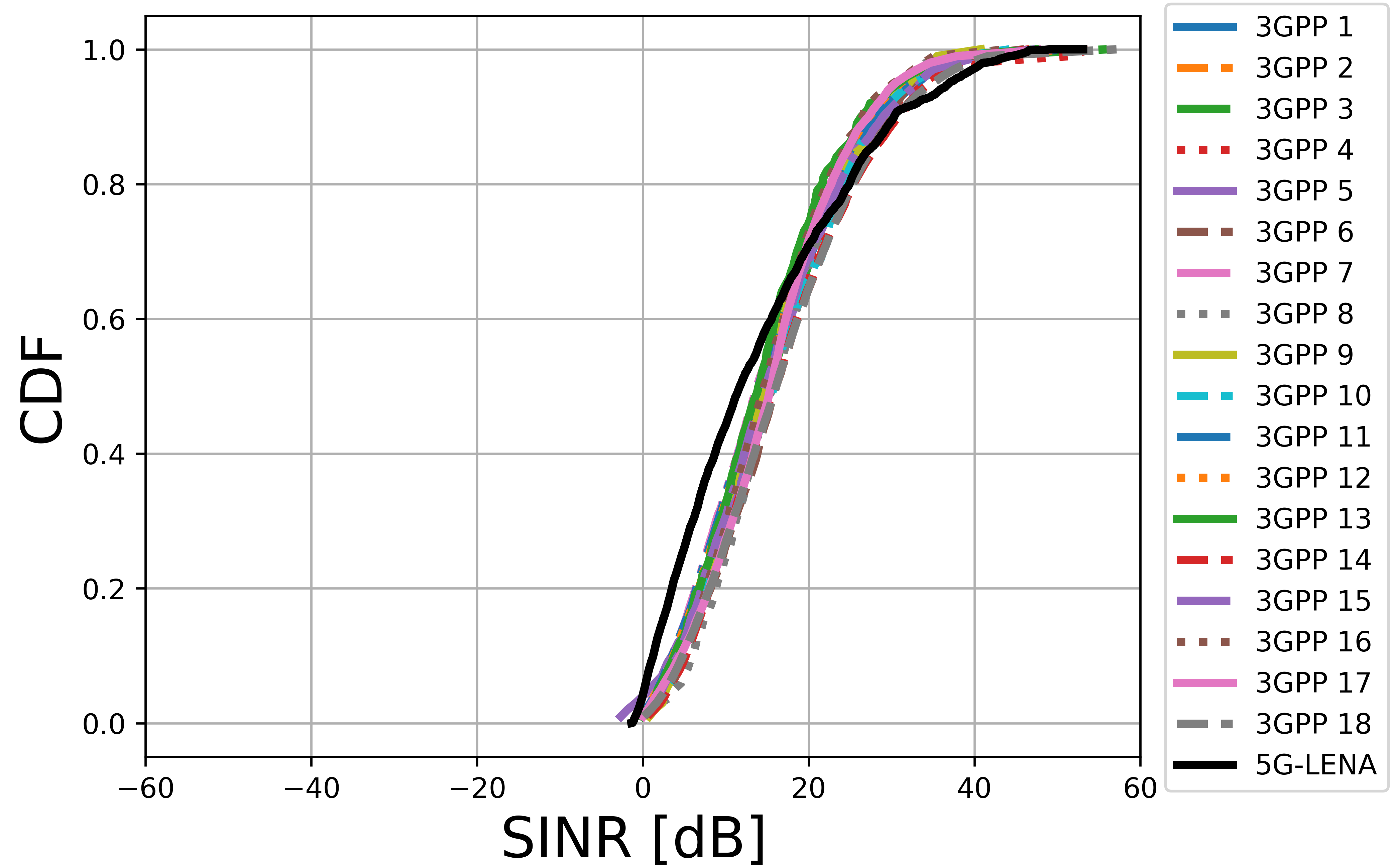}
\centering
\caption{Dense Urban Configuration A - Downlink Geometry}
\label{fig:SinrDenseA}
\end{figure}

Furthermore, results for Dense Urban-eMBB are shown in Figures~\ref{fig:CgDenseA} and~\ref{fig:SinrDenseA}, for the Coupling Gain and Downlink Geometry (SINR), respectively. In this case, the 5G-LENA simulator calibrates quite well with respect to the SINR compared to the results of the benchmark radio interface technologies. However, there is a small difference in the Coupling Gain due to the fact that the 5G-LENA traces for the Coupling Gain do not include the analog beamforming gains. So, the difference is justified because of the missing beamforming gain in our Coupling Gain computation in 5G-LENA (note that the Urban scenarios do not include beamforming gain as per their definition, see Table~\ref{table:Ruraltable}).
Despite this difference, it can be seen that the 5G-LENA simulator calibrates well with respect to the 3GPP SINR specifications, since for the SINR trace we properly account for all the factors.

Based on the above results, it is proven that the open-source 5G-LENA system-level simulator can be exploited by the research community for the study and testing of existing and/or new features providing reliable results that are analogous to that of the 3GPP industrial simulators.

\section{Conclusions}
\label{sec:Conclusions}
In this paper, we have confirmed the validity and correct functionality of the open-source 5G-LENA (NR) simulator through the performance evaluation against 3GPP industrial simulators. Test environments are defined in the ITU-R IMT-2020 Report, as well as the minimum technical performance requirements. The calibration process followed in this paper has been carried out in two steps. First, 5G-LENA REM feature has been used as tool to understand the radio environment, such as the signal propagation, the antenna configuration and the network interference, proving the importance of having available such a tool in system-level simulators. Second, end-to-end simulations have been performed and results have been compared against the reference industrial simulators based on the 3GPP defined KPIs. The resemblance of the results with that provided by 3GPP proves that the simulator is analogous to industrial or private ones and can be used by the research community for the development, study and evaluation of existing and/or new features.

\section*{Acknowledgment}
This work was partially funded by Meta-Facebook and Spanish MINECO grant TSI-063000-2021-56/TSI-063000-2021-57 (6G-BLUR). The authors would like to thank Xiaodi Zhang for the fruitful discussions and help.

\bibliography{main}

\begin{thebibliography}{30}
\expandafter\ifx\csname natexlab\endcsname\relax\def\natexlab#1{#1}\fi
\providecommand{\url}[1]{\texttt{#1}}
\providecommand{\href}[2]{#2}
\providecommand{\path}[1]{#1}
\providecommand{\DOIprefix}{doi:}
\providecommand{\ArXivprefix}{arXiv:}
\providecommand{\URLprefix}{URL: }
\providecommand{\Pubmedprefix}{pmid:}
\providecommand{\doi}[1]{\href{http://dx.doi.org/#1}{\path{#1}}}
\providecommand{\Pubmed}[1]{\href{pmid:#1}{\path{#1}}}
\providecommand{\bibinfo}[2]{#2}
\ifx\xfnm\relax \def\xfnm[#1]{\unskip,\space#1}\fi
\bibitem[{3GPP(2021)}]{38300}
\bibinfo{author}{3GPP}, \bibinfo{title}{{NR and NG-RAN Overall Description}},
  \bibinfo{howpublished}{TS 38.300 (Rel. 15), v16.7.0}, \bibinfo{year}{2021}.
\bibitem[{Parkvall et~al.(2017)}]{parkvall:17}
\bibinfo{author}{S.~Parkvall}, et~al.,
\newblock \bibinfo{title}{{NR: the new 5G radio access technology}},
\newblock \bibinfo{journal}{IEEE Commun. Standards Mag.} \bibinfo{volume}{1}
  (\bibinfo{year}{2017}) \bibinfo{pages}{24--30}.
\bibitem[{Pi and Khan(2011)}]{pi:11}
\bibinfo{author}{Z.~Pi}, \bibinfo{author}{F.~Khan},
\newblock \bibinfo{title}{{An introduction to millimeter-wave mobile broadband
  systems}},
\newblock \bibinfo{journal}{IEEE Commun. Magazine} \bibinfo{volume}{49}
  (\bibinfo{year}{2011}) \bibinfo{pages}{101–--107}.
\bibitem[{{Peng} et~al.(2016){Peng}, {Sun}, {Li}, {Mao}, and {Wang}}]{Peng16}
\bibinfo{author}{M.~{Peng}}, \bibinfo{author}{Y.~{Sun}},
  \bibinfo{author}{X.~{Li}}, \bibinfo{author}{Z.~{Mao}},
  \bibinfo{author}{C.~{Wang}},
\newblock \bibinfo{title}{Recent advances in cloud radio access networks:
  System architectures, key techniques, and open issues},
\newblock \bibinfo{journal}{IEEE Communications Surveys Tutorials}
  \bibinfo{volume}{18} (\bibinfo{year}{2016}) \bibinfo{pages}{2282--2308}.
\bibitem[{Gkonis et~al.(2020)Gkonis, Trakadas, and
  Kaklamani}]{electronics9030468}
\bibinfo{author}{P.~K. Gkonis}, \bibinfo{author}{P.~T. Trakadas},
  \bibinfo{author}{D.~I. Kaklamani},
\newblock \bibinfo{title}{A comprehensive study on simulation techniques for 5g
  networks: State of the art results, analysis, and future challenges},
\newblock \bibinfo{journal}{Electronics} \bibinfo{volume}{9}
  (\bibinfo{year}{2020}).
\bibitem[{Sun et~al.(2017)Sun, MacCartney, and Rappaport}]{NYUSIM}
\bibinfo{author}{S.~Sun}, \bibinfo{author}{G.~R. MacCartney},
  \bibinfo{author}{T.~S. Rappaport},
\newblock \bibinfo{title}{A novel millimeter-wave channel simulator and
  applications for {5G} wireless communications},
\newblock in: \bibinfo{booktitle}{2017 IEEE International Conference on
  Communications (ICC)}, \bibinfo{year}{2017}, pp. \bibinfo{pages}{1--7}.
\bibitem[{M{\"u}ller et~al.(2018)M{\"u}ller, Ademaj, Dittrich, Fastenbauer,
  Ramos~Elbal, Nabavi, Nagel, Schwarz, and Rupp}]{vienna5g}
\bibinfo{author}{M.~K. M{\"u}ller}, \bibinfo{author}{F.~Ademaj},
  \bibinfo{author}{T.~Dittrich}, \bibinfo{author}{A.~Fastenbauer},
  \bibinfo{author}{B.~Ramos~Elbal}, \bibinfo{author}{A.~Nabavi},
  \bibinfo{author}{L.~Nagel}, \bibinfo{author}{S.~Schwarz},
  \bibinfo{author}{M.~Rupp},
\newblock \bibinfo{title}{Flexible multi-node simulation of cellular mobile
  communications: the {V}ienna {5G} {S}ystem {L}evel {S}imulator},
\newblock \bibinfo{journal}{EURASIP Journal on Wireless Communications and
  Networking} \bibinfo{volume}{2018} (\bibinfo{year}{2018})
  \bibinfo{pages}{227}.
\bibitem[{Jao et~al.(2018)Jao, Wang, Yeh, Tsai, Lo, Chen, Pao, and
  Sheen}]{wiseSim}
\bibinfo{author}{C.-K. Jao}, \bibinfo{author}{C.-Y. Wang},
  \bibinfo{author}{T.-Y. Yeh}, \bibinfo{author}{C.-C. Tsai},
  \bibinfo{author}{L.-C. Lo}, \bibinfo{author}{J.-H. Chen},
  \bibinfo{author}{W.-C. Pao}, \bibinfo{author}{W.-H. Sheen},
\newblock \bibinfo{title}{{WiSE}: A system-level simulator for {5G} mobile
  networks},
\newblock \bibinfo{journal}{IEEE Wireless Communications} \bibinfo{volume}{25}
  (\bibinfo{year}{2018}) \bibinfo{pages}{4--7}.
\bibitem[{Dominguez-Bolano et~al.(2016)Dominguez-Bolano, Rodriguez-Pineiro,
  Garcia-Naya, and Castedo}]{GTECsim}
\bibinfo{author}{T.~Dominguez-Bolano}, \bibinfo{author}{J.~Rodriguez-Pineiro},
  \bibinfo{author}{J.~A. Garcia-Naya}, \bibinfo{author}{L.~Castedo},
\newblock \bibinfo{title}{The {GTEC 5G} link-level simulator},
\newblock in: \bibinfo{booktitle}{2016 1st International Workshop on Link- and
  System Level Simulations (IWSLS)}, \bibinfo{year}{2016}, pp.
  \bibinfo{pages}{1--6}.
\bibitem[{mat(2021)}]{matlab5g}
\bibinfo{title}{{5G} toolbox}, \bibinfo{howpublished}{Available at {\tt
  https://www.mathworks.com/products/5g.html}}, \bibinfo{year}{2021}.
  \bibinfo{note}{Accessed: 23-03-22}.
\bibitem[{Nardini et~al.(2020)Nardini, Stea, Virdis, and Sabella}]{simu5G}
\bibinfo{author}{G.~Nardini}, \bibinfo{author}{G.~Stea},
  \bibinfo{author}{A.~Virdis}, \bibinfo{author}{D.~Sabella},
\newblock \bibinfo{title}{Simu5g: A system-level simulator for 5g networks},
\newblock in: \bibinfo{booktitle}{Proceedings of the 10th International
  Conference on Simulation and Modeling Methodologies, Technologies and
  Applications - SIMULTECH}, \bibinfo{organization}{INSTICC},
  \bibinfo{publisher}{SciTePress}, \bibinfo{year}{2020}, pp.
  \bibinfo{pages}{68--80}.
\bibitem[{Martiradonna et~al.(2020)Martiradonna, Grassi, Piro, and
  Boggia}]{5GairSim}
\bibinfo{author}{S.~Martiradonna}, \bibinfo{author}{A.~Grassi},
  \bibinfo{author}{G.~Piro}, \bibinfo{author}{G.~Boggia},
\newblock \bibinfo{title}{5g-air-simulator: An open-source tool modeling the 5g
  air interface},
\newblock in: \bibinfo{booktitle}{Computer Networks}, volume
  \bibinfo{volume}{173}, \bibinfo{year}{2020}, p. \bibinfo{pages}{107151}.
\bibitem[{Mezzavilla et~al.(2018)Mezzavilla, Zhang, Polese, Ford, Dutta,
  Rangan, and Zorzi}]{nyummwave}
\bibinfo{author}{M.~Mezzavilla}, \bibinfo{author}{M.~Zhang},
  \bibinfo{author}{M.~Polese}, \bibinfo{author}{R.~Ford},
  \bibinfo{author}{S.~Dutta}, \bibinfo{author}{S.~Rangan},
  \bibinfo{author}{M.~Zorzi},
\newblock \bibinfo{title}{{End-to-End Simulation of 5G mmWave Networks}},
\newblock \bibinfo{journal}{IEEE Communications Surveys Tutorials}
  \bibinfo{volume}{20} (\bibinfo{year}{2018}) \bibinfo{pages}{2237--2263}.
\bibitem[{Patriciello et~al.(2019)Patriciello, Lagen, Bojovic, and
  Giupponi}]{PATRICIELLO2019101933}
\bibinfo{author}{N.~Patriciello}, \bibinfo{author}{S.~Lagen},
  \bibinfo{author}{B.~Bojovic}, \bibinfo{author}{L.~Giupponi},
\newblock \bibinfo{title}{{An E2E Simulator for 5G NR Networks}},
\newblock \bibinfo{journal}{Simulation Modelling Practice and Theory}
  \bibinfo{volume}{96} (\bibinfo{year}{2019}) \bibinfo{pages}{101933}.
\bibitem[{{ITU-R}(2019)}]{IMT-2020}
\bibinfo{author}{{ITU-R}}, \bibinfo{title}{{Submission, evaluation process and
  consensus building for IMT-2020}}, \bibinfo{howpublished}{ITU-R
  IMT-2020/2-E}, \bibinfo{year}{2019}.
\bibitem[{Koutlia et~al.(2021)Koutlia, Bojovi\'{c}, Lag\'{e}n, and
  Giupponi}]{rem3}
\bibinfo{author}{K.~Koutlia}, \bibinfo{author}{B.~Bojovi\'{c}},
  \bibinfo{author}{S.~Lag\'{e}n}, \bibinfo{author}{L.~Giupponi},
\newblock \bibinfo{title}{Novel radio environment map for the ns-3 nr
  simulator},
\newblock in: \bibinfo{booktitle}{Proceedings of the Workshop on Ns-3}, WNS3
  '21, \bibinfo{address}{New York, NY, USA}, \bibinfo{year}{2021}, p.
  \bibinfo{pages}{41–48}.
\bibitem[{Perez-Romero et~al.(2015)Perez-Romero, Zalonis, Boukhatem, Kliks,
  Koutlia, Dimitriou, and Kurda}]{rem2}
\bibinfo{author}{J.~Perez-Romero}, \bibinfo{author}{A.~Zalonis},
  \bibinfo{author}{L.~Boukhatem}, \bibinfo{author}{A.~Kliks},
  \bibinfo{author}{K.~Koutlia}, \bibinfo{author}{N.~Dimitriou},
  \bibinfo{author}{R.~Kurda},
\newblock \bibinfo{title}{On the use of radio environment maps for interference
  management in heterogeneous networks},
\newblock \bibinfo{journal}{IEEE Communications Magazine} \bibinfo{volume}{53}
  (\bibinfo{year}{2015}) \bibinfo{pages}{184--191}.
\bibitem[{Henderson et~al.(2008)Henderson, Lacage, Riley, Dowell, and
  Kopena}]{ns3}
\bibinfo{author}{T.~R. Henderson}, \bibinfo{author}{M.~Lacage},
  \bibinfo{author}{G.~F. Riley}, \bibinfo{author}{C.~Dowell},
  \bibinfo{author}{J.~Kopena},
\newblock \bibinfo{title}{{Network simulations with the ns-3 simulator}},
\newblock \bibinfo{journal}{SIGCOMM demonstration} \bibinfo{volume}{14}
  (\bibinfo{year}{2008}).
\bibitem[{5gL(2019)}]{5gLena}
\bibinfo{title}{{5G-LENA}}, \bibinfo{howpublished}{Available at {\tt
  https://5g-lena.cttc.es/}}, \bibinfo{year}{2019}.
\bibitem[{Baldo et~al.(2011)Baldo, Miozzo, Requena-Esteso, and
  Nin-Guerrero}]{BaldoLena}
\bibinfo{author}{N.~Baldo}, \bibinfo{author}{M.~Miozzo},
  \bibinfo{author}{M.~Requena-Esteso}, \bibinfo{author}{J.~Nin-Guerrero},
\newblock \bibinfo{title}{An open source product-oriented lte network simulator
  based on ns-3},
\newblock in: \bibinfo{booktitle}{Proceedings of the 14th ACM International
  Conference on Modeling, Analysis and Simulation of Wireless and Mobile
  Systems}, MSWiM '11, \bibinfo{publisher}{ACM}, \bibinfo{address}{New York,
  NY, USA}, \bibinfo{year}{2011}, pp. \bibinfo{pages}{293--298}.
\bibitem[{Marinescu et~al.(2017)Marinescu, Macaluso, and DaSilva}]{marinescu}
\bibinfo{author}{A.~Marinescu}, \bibinfo{author}{I.~Macaluso},
  \bibinfo{author}{L.~A. DaSilva},
\newblock \bibinfo{title}{System level evaluation and validation of the ns-3
  lte module in 3gpp reference scenarios},
\newblock in: \bibinfo{booktitle}{Proceedings of the 13th ACM Symposium on QoS
  and Security for Wireless and Mobile Networks}, \bibinfo{year}{2017}, p.
  \bibinfo{pages}{59–64}.
\bibitem[{{ITU-R}(2017{\natexlab{a}})}]{M.2410}
\bibinfo{author}{{ITU-R}}, \bibinfo{title}{{Minimum Requirements related to
  Technical Performance for IMT-2020 Radio Interface(s)}},
  \bibinfo{howpublished}{Report M.2410-0}, \bibinfo{year}{November
  2017}{\natexlab{a}}.
\bibitem[{{ITU-R}(2017{\natexlab{b}})}]{M.2411}
\bibinfo{author}{{ITU-R}}, \bibinfo{title}{{Requirements, Evaluation Criteria
  and Submission Templates for the Development of IMT-2020}},
  \bibinfo{howpublished}{Report M.2411-0}, \bibinfo{year}{November
  2017}{\natexlab{b}}.
\bibitem[{{ITU-R}(2017{\natexlab{c}})}]{M.2412}
\bibinfo{author}{{ITU-R}}, \bibinfo{title}{{Guidelines for Evaluation of Radio
  Interface Technologies for IMT-2020}}, \bibinfo{howpublished}{Report
  M.2412-0}, \bibinfo{year}{October 2017}{\natexlab{c}}.
\bibitem[{{Huawei}(2018)}]{RP180524}
\bibinfo{author}{{Huawei}}, \bibinfo{title}{{Summary of Calibration Results for
  IMT-2020 Self Evaluation}}, \bibinfo{howpublished}{3GPP TSG RAN Meeting 79,
  RP-180524}, \bibinfo{year}{2018}.
\bibitem[{{3GPP}(2020)}]{TR38901}
\bibinfo{author}{{3GPP}}, \bibinfo{title}{{Study on Channel Model for
  Frequencies from 0.5 to 100 GHz}}, \bibinfo{howpublished}{TR 38.901 (Rel.
  15), v16.1.0}, \bibinfo{year}{2020}.
\bibitem[{Zugno et~al.(2020)Zugno, Polese, Patriciello, Bojovic, Lagen, and
  Zorzi}]{tommaso:20}
\bibinfo{author}{T.~Zugno}, \bibinfo{author}{M.~Polese},
  \bibinfo{author}{N.~Patriciello}, \bibinfo{author}{B.~Bojovic},
  \bibinfo{author}{S.~Lagen}, \bibinfo{author}{M.~Zorzi},
\newblock \bibinfo{title}{Implementation of a spatial channel model for ns-3},
\newblock in: \bibinfo{booktitle}{Proceedings of the 2020 Workshop on ns-3},
  \bibinfo{address}{Gaithersburg, MD, USA}, \bibinfo{year}{2020}, p.
  \bibinfo{pages}{49–56}.
\bibitem[{Yu et~al.(2018)Yu, Dietrich, and Pauli}]{Nomor}
\bibinfo{author}{L.~Yu}, \bibinfo{author}{C.~Dietrich},
  \bibinfo{author}{V.~Pauli},
\newblock \bibinfo{title}{Imt-2020 evaluation : Calibration of nomor ’ s
  system level simulator},
\newblock \bibinfo{year}{2018}.
\bibitem[{ns3(2022{\natexlab{a}})}]{ns3devCttc}
\bibinfo{howpublished}{Available at {\tt https://gitlab.com/cttc-lena/nr}},
  \bibinfo{year}{2022}{\natexlab{a}}.
\bibitem[{ns3(2022{\natexlab{b}})}]{ns3Calibration}
\bibinfo{howpublished}{Available at {\tt
  https://gitlab.com/cttc-lena/ns-3-dev/-/tree/calibration}},
  \bibinfo{year}{2022}{\natexlab{b}}.

\end{thebibliography}

\end{document}